\documentclass[acmtog,balance = false]{acmart}

\usepackage{booktabs} 
\usepackage{amsmath}
\usepackage{amsbsy}

\acmPrice{15.00}

\setcitestyle{square}

\setcopyright{acmcopyright}\acmJournal{TOG}
\acmYear{2021}\acmVolume{40}\acmNumber{4}\acmArticle{122}\acmMonth{8} \acmDOI{10.1145/3450626.3459769}

\newcommand{\tetl}{\text{tetrahedral}} 
\newcommand{\tet}{\text{tetrahedron}}
\newcommand{\tets}{\text{tetrahedrons}}
\newcommand{\tissueit}{\textit{tissue}} 
\newcommand{\tissue}{tissue} 
\renewcommand{\c}{\vc{c}}
\DeclareMathOperator*{\argmax}{argmax}

\usepackage{microtype} 
\usepackage{xspace}
\usepackage{enumitem}

\usepackage{soul}
\usepackage{breakcites}
\usepackage{ellipsis} 
\usepackage[mathletters]{ucs}
\usepackage[utf8x]{inputenc}
\usepackage{mathtools}
\usepackage{amsbsy}
\usepackage{upgreek}
\usepackage{dblfloatfix}

\definecolor{lightbluishgrey}{rgb}{0.76078,0.88235,0.92157}

\newcommand{\rev}[1]{#1}
\newcommand{\rem}[1]{}
\newcommand{\revtwo}[1]{#1}

\usepackage{layouts}
\usepackage{wrapfig}

\newcommand{\reffig}[1] {Fig.~\ref{fig:#1}}

\makeatletter
\def\reffig{\@ifnextchar[{\@myreffigloc}{\@myreffignoloc}}
\def\@myreffigloc[#1]#2{Fig.~\ref{fig:#2}, \emph{#1}}
\def\@myreffignoloc#1{Fig.~\ref{fig:#1}}
\makeatother

\let\mat = \mathbf
\usepackage{dsfont}

\newcommand{\R}{\mathbb{R}}
\newcommand{\Q}{\mathbf{Q}}
\newcommand{\vc}[1]{\mathbf{#1}}
\renewcommand{\C}{\mat{C}}

\renewcommand{\d}{\mathrm{d}}

\newcommand{\e}{\vc{e}}
\newcommand{\f}{\vc{f}}
\newcommand{\g}{\vc{g}}

\newcommand{\n}{\hat{\vc{n}}}

\renewcommand{\t}{\vc{t}}
\renewcommand{\u}{\vc{u}}

\newcommand{\A}{\mat{A}}
\newcommand{\B}{\mat{B}}
\renewcommand{\C}{\mat{C}}

\newcommand{\E}{\mat{E}}

\renewcommand{\G}{\mat{G}}

\renewcommand{\L}{\mat{L}}
\newcommand{\M}{\mat{M}}

\newcommand{\N}{\mat{N}}

\newcommand{\T}{\mat{T}}

\newcommand{\V}{\mat{V}}

\newcommand{\W}{\mat{W}}
\newcommand{\X}{\mat{X}}

\usepackage{siunitx}
\usepackage{graphicx}

\usepackage{wrapfig}

\makeatletter
\let\save@mathaccent\mathaccent
\newcommand*\if@single[3]{%
  \setbox0\hbox{${\mathaccent"0362{#1}}^H$}%
  \setbox2\hbox{${\mathaccent"0362{\kern0pt#1}}^H$}%
  \ifdim\ht0=\ht2 #3\else #2\fi
  }
\newcommand*\rel@kern[1]{\kern#1\dimexpr\macc@kerna}
\newcommand*\widebar[1]{\@ifnextchar^{{\wide@bar{#1}{0}}}{\wide@bar{#1}{1}}}
\newcommand*\wide@bar[2]{\if@single{#1}{\wide@bar@{#1}{#2}{1}}{\wide@bar@{#1}{#2}{2}}}
\newcommand*\wide@bar@[3]{%
  \begingroup
  \def\mathaccent##1##2{%
    \let\mathaccent\save@mathaccent
    \if#32 \let\macc@nucleus\first@char \fi
    \setbox\z@\hbox{$\macc@style{\macc@nucleus}_{}$}%
    \setbox\tw@\hbox{$\macc@style{\macc@nucleus}{}_{}$}%
    \dimen@\wd\tw@
    \advance\dimen@-\wd\z@
    \divide\dimen@ 3
    \@tempdima\wd\tw@
    \advance\@tempdima-\scriptspace
    \divide\@tempdima 10
    \advance\dimen@-\@tempdima
    \ifdim\dimen@>\z@ \dimen@0pt\fi
    \rel@kern{0.6}\kern-\dimen@
    \if#31
      \overline{\rel@kern{-0.6}\kern\dimen@\macc@nucleus\rel@kern{0.4}\kern\dimen@}%
      \advance\dimen@0.4\dimexpr\macc@kerna
      \let\final@kern#2%
      \ifdim\dimen@<\z@ \let\final@kern1\fi
      \if\final@kern1 \kern-\dimen@\fi
    \else
      \overline{\rel@kern{-0.6}\kern\dimen@#1}%
    \fi
  }%
  \macc@depth\@ne
  \let\math@bgroup\@empty \let\math@egroup\macc@set@skewchar
  \mathsurround\z@ \frozen@everymath{\mathgroup\macc@group\relax}%
  \macc@set@skewchar\relax
  \let\mathaccentV\macc@nested@a
  \if#31
    \macc@nested@a\relax111{#1}%
  \else
    \def\gobble@till@marker##1\endmarker{}%
    \futurelet\first@char\gobble@till@marker#1\endmarker
    \ifcat\noexpand\first@char A\else
      \def\first@char{}%
    \fi
    \macc@nested@a\relax111{\first@char}%
  \fi
  \endgroup
}
\makeatother

\usepackage[ruled,vlined]{algorithm2e}
\usepackage{etoolbox}
\usepackage{array}

\newcommand{\figs}{}
\def\figs/{figs/}

\DeclareMathOperator*{\argmin}{argmin}

\usepackage[linewidth=1pt]{mdframed}

\newcounter{question}
\newcommand{\Question}[1]%
{%
  \stepcounter{question}%
  \textbf{\textcolor[rgb]{0.9,0.5,0.3}{\textsc{Question \thequestion}:}} %
  \emph{#1}%
}
\newcommand{\Hypothesis}[1]%
{%
  \textbf{\textcolor[rgb]{0.3,0.3,0.7}{\textsc{Hypothesis \thequestion}:}} {#1}
}
\newcommand{\Answer}[1]%
{%
  \textbf{\textcolor[rgb]{0.3,0.7,0.3}{\textsc{Answer \thequestion}:}} {#1}
}

\let\[\equation
\let\]\endequation

\usepackage{bbm}

\usepackage{lipsum}

\graphicspath{{images/}}



\begin{document}

\title{Interactive Modelling of Volumetric Musculoskeletal Anatomy} 

\author{Rinat Abdrashitov}
\affiliation{
  \institution{University of Toronto}
  \city{Toronto}
  \country{Canada}}
\email{rinat@dgp.toronto.edu}

\author{Seungbae Bang}
\affiliation{
  \institution{University of Toronto}
  \city{Toronto}
  \country{Canada}}
\email{seungbae@cs.toronto.edu}

\author{David Levin}
\affiliation{
  \institution{University of Toronto}
  \country{Canada}}
\email{diwlevin@cs.toronto.edu}

\author{Karan Singh}
\affiliation{
  \institution{University of Toronto}
  \city{Toronto}
  \country{Canada}}
\email{karan@dgp.toronto.edu}

\author{Alec Jacobson}
\affiliation{
  \institution{University of Toronto}
  \city{Toronto}
  \country{Canada}}
\email{jacobson@cs.toronto.edu}

\begin{abstract}
    We present a new approach for modelling \rem{animation ready} musculoskeletal anatomy. 
    Unlike previous methods, we do not model individual muscle shapes as geometric 
    primitives (polygonal meshes, NURBS etc.). Instead, we adopt a volumetric 
    segmentation approach where every point in our volume is assigned to a muscle, 
    fat, or bone tissue. We provide an interactive modelling tool where the user 
    controls the segmentation via muscle curves and we visualize the muscle shapes 
    using volumetric rendering. Muscle curves enable intuitive yet powerful control 
    over the muscle shapes. This representation allows us to automatically handle 
    intersections between different tissues (muscle-muscle, muscle-bone, and muscle-skin) 
    \rev{during the modelling} and automates computation of muscle fiber fields. 
    We further introduce a novel algorithm for converting the volumetric muscle 
    representation into tetrahedral or surface geometry for use in downstream 
    tasks. \rem{like simulation, skinning, and visualization.} Additionally, we 
    introduce an interactive skeleton authoring tool that allows the users to 
    create skeletal anatomy starting from only a skin mesh using a library 
    of bone parts.
\end{abstract}

\begin{teaserfigure}
    \includegraphics[width=\textwidth, keepaspectratio=true]{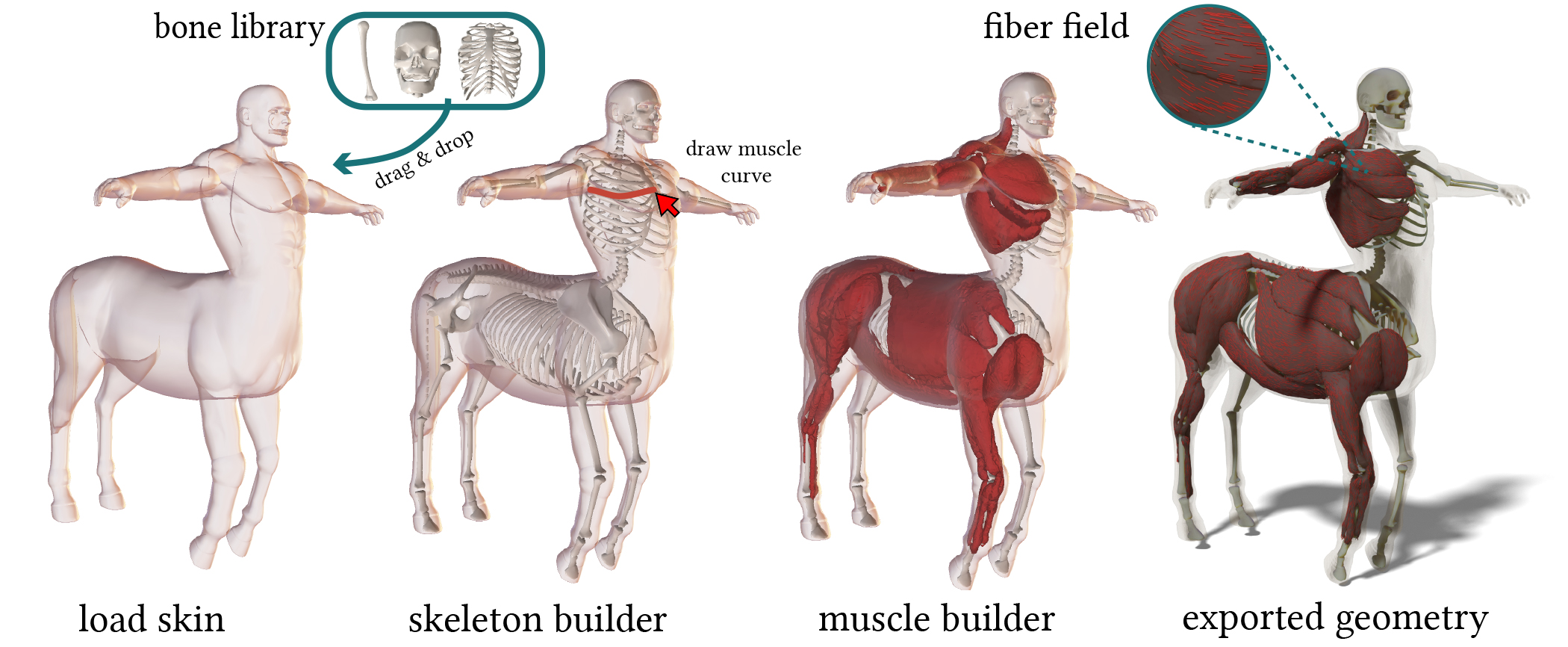}
    \caption{Given a skin surface mesh, a library of bone parts is used to quickly 
    create a skeleton in our skeleton builder tool. The user then draws curves to 
    generate the muscle shapes which are visualized using a volume rendering. Once 
    all muscles are created, we can export the geometry of the muscles, automatically 
    compute fiber fields and use the result in downstream applications. \rem{like simulation, 
    skinning, and visualization.} Centaur model is part of the TOSCA dataset 
    \cite{bronstein2008numerical}. Horse bone models were obtained from 
    \url{https://3dassets.store/}. Used under permission.}
    \label{fig:teaser}
\end{teaserfigure}

\maketitle
\section{Introduction}
Digital characters are a driving force in the entertainment industry 
allowing artists to tell stories limited only by their imagination. 
A lot of effort goes into reaching a point where digital characters are 
indistinguishable from the real ones. Characters are 
often modeled by only considering their skin \cite{jacobson2014skinning}, 
disregarding underlying 
volumetric muscle, fat, and bone structure. 
Animating physically realistic effects
like muscle bulging, skin sliding, wrinkles, and volume preservation, without 
an explicit musculoskeletal structure is challenging and, and requires skilled 
and tedious manual effort, to achieve high quality results. 
Therefore, a truly accurate portrayal of digital characters requires the creation 
of biologically representative musculoskeletal anatomy. 

Different solutions that allow artists to automate tedious manual tasks like the 
creation of the skin \cite{yoshiyasu2014conformal}, hair \cite{saito20183d}, 
rigs \cite{xu2020rignet} and others have been explored over the years. However, 
user-friendly solutions to the problem of creating a musculoskeletal structure that 
is suitable for character modelling and animation are relatively unexplored. The current 
solutions either require artists to model every muscle using sculpting software, 
go through tedious parameter tweaking of geometric primitives, or create a detailed 
template for retargeting to new geometries. These methods make it hard to produce 
complex intersection-free muscle shapes that conform to the skin surface. Additionally, 
defining muscle fiber directions requires users to manually specify attachment 
points for every muscle. Making incremental changes using these approaches is 
tedious and hinders the fast exploration of character design.

\rem{Anatomical models are typically constructed from the inside out. 
Internal tissues are modeled first and a surface skin is attached to, or generated 
from, the internal structure.} We propose an interactive modelling tool, that adopts 
the \textit{outside–in} \cite{pratscher2005outside} approach and enables the creation 
of a musculoskeletal system starting from a skin mesh. The user starts by arranging 
the skeleton from pre-existing templates of bones. Then the user simply
sketches curves inside a volume constrained by the skin and our system automatically 
infers the muscle shapes. Inspired by \citet{orzan2008diffusion}, 
we utilize a diffusion process to segment the volume into muscle and fat tissues
based on the user-created curve network. The resulting muscles are intersection 
free and conform to the skin geometry. We show how to utilize volume rendering
to visualize the muscle shapes and hence avoid the need for explicit meshing every 
time the user edits a curve. We further propose an algorithm for extracting the 
manifold muscle meshes from our volumetric segmentation for the use in downstream 
tasks. \rem{like simulation, skinning, or visualization.} To the best of our knowledge, 
our system is the first user-friendly interactive modelling tool, capable of 
creating \rem{an animation ready} \rev{intersection free geometry for the} 
musculoskeletal system.
\section{Background and Related Work}
We provide a brief background on muscle anatomy followed by review of related work
categorized by muscle representations in physically-based animation; skinning methods 
that geometrically attempt to “emulate” the physics of muscle deformation; anatomic 
templates to aid character modelling, setup and transfer; and interactive interfaces 
for volumetric and character modelling.
\subsection{Musculoskeletal Anatomy:}
\begin{figure}
    \includegraphics[width=\columnwidth]{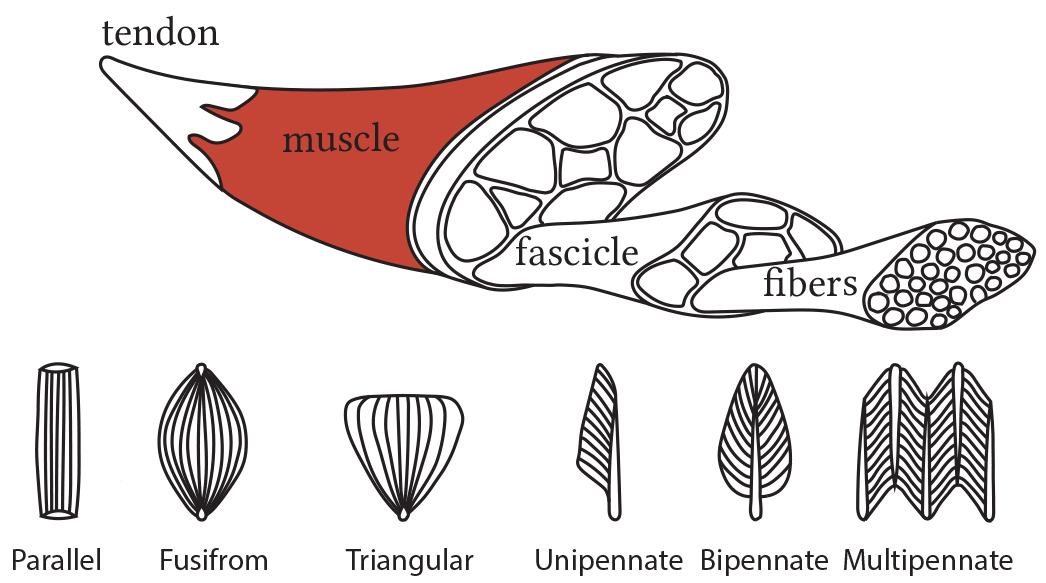}
    \caption{Muscle structure.
    Image courtesy of \cite{lee2010survey}}
    \label{fig:muscle_structure} 
\end{figure}
Muscle is a soft tissue (of type skeletal, cardiac or smooth), whose function is 
to produce force and motion. A large body of research in Computer Graphics, 
biomechanics and robotics is focused on studying the physiological properties and 
and function of skeletal muscles \cite{scheepers97, ng97}. For the rest of the 
paper, we will refer to "skeletal muscles" simply as "muscles". Internally, the 
muscle is composed of numerous muscle \textit{fiber} bundles, called \textit{fascicles}. 
Large muscles, such as the biceps brachii or the 
sartorius have \textit{fascicles} arranged parallel to one another 
along the length of the muscle.  
Other muscles exhibit fascicles with a \textit{pennation} angle, between their tendinous 
attachments and the longitudinal axis of the muscle 
(Fig \ref{fig:muscle_structure} bottom).
Skeletal muscle is 
anchored by tendons to bone. Tendons transmit forces produced by the attached 
muscle to the bone, enabling locomotion and maintaining posture (Fig. 
\ref{fig:muscle_structure} top).  
We refer the reader to the survey by \citet{lee2010survey} which provides a 
thorough overview of modelling and simulation of skeletal muscles.

\subsection{Surface-based muscle primitives:}
Musculoskeletal primitives for geometric character skinning is at least three 
decades old \cite{chadwick89}. Early research has explored the formulation of 
muscles as collections of ellipsoids \cite{singh95,scheepers97,pratscher2005outside}, 
generalized cylinders \cite{wilhelms97,simmons02},  polygon meshes \cite{albrecht03}, 
extruded parametric curves \cite{helpinghand},  NURBS \cite{mayamuscle}, and 
implicit models \cite{roussellet2018dynamic}. These surface-based muscle primitives 
typically serve as proxy geometry to bind and geometrically deform a geometric 
skin. While these primitives can be imbued with simplified muscle dynamics, 
they are ill-suited to general purpose anatomic simulation \cite{ziva}. Our 
representation, based on muscle curves and anisotropically induced muscle 
volumes, provides the high-level geometric control of muscle shape and skin 
deformation of these surface-based primitives,  but can also produce various 
muscle shapes (parallel, convergent, pennate) and automate the computation 
of fiber fields. Being an inherently volume-based representation, it 
is also well-suited to handle general muscle-muscle, muscle-bone, muscle-skin 
intersections  and  muscle fiber bundle computations. \rem{necessary for the physcial 
simulation of muscloskeletal anatomy.}



\subsection{Volume-based muscle primitives:}
Physically-based simulation of the skin layered over volumetric muscle 
primitives \cite{li2013thin} is a \rev{desirable} solution to producing the
subtle details of skin motion \cite{ziva,wetatissue}. 
The initial musuloskeletal setup of a character as comprised of skin, fat, muscle 
and bone, in a simulation pipeline is tedious and 
requires multiple iterations of laboriously rebuilding hand-crafted bone and 
muscle geometry \cite{deepak}, to elicit the desired simulation behavior from 
the musculoskeletal anatomy.  These sculpted geometries then need to be 
processed to resolve intersections, and define fiber fields to support anisotropic 
muscle contraction. 
While MRI/CT scan data can aid the reconstruction of accurate live anatomy 
\cite{teran2005creating, jacobs2016build}, such data must be artist-imagined 
for fictional characters. 
Muscles have also been built by physically simulating inflatable 3D 
patches defined by a user on a character's skin \cite{turchet2017physically},
or as parametric solid volumes \cite{ng97}, but these muscles tend to leave  
undesirable gaps between muscles, bones and other internal structures.
\citet{angles2019viper} models a muscle as a bundle of position-based rods 
augmented with isotropic scale to enable simulation of volumetric effects. 
However, their rod-based representation requires users to either manually create 
bundles or acquire pre-existing tetrahedral geometry of muscles which is then 
automatically converted to their representation. 
\citet{yu2020repulsive} introduces an efficient algorithm for (self)-repulsion of  
space curves that can be used to design biologically-inspired
curve networks such as muscle fibers. However, their optimization-based approach 
is not suitable for interactive modelling of a large number of muscles. 



\subsection{Anatomic Templates:}
The first semi-automatic method for creating  anatomical structures, 
such as bones, muscles, viscera, and fat tissues was proposed by \citet{ali2013anatomy}. 
Their method can be seen as a partial registration process, where skin surfaces 
are first registered based on the data, and the interior tissues are estimated 
using interpolation and anatomical rules. \citet{saito2015computational} 
create a wide range of human body shapes from a single input 3D anatomy template 
by simulating biological processes responsible for human body growth. 
\citet{kadlevcek2016reconstructing} use a set of 3D scans of an
actor in various poses to compute subject-specific and pose-dependent parameters 
of an anatomical template model, to explain the captured 3D scans as closely as
possible. \rev{Our method is complementary to these approaches and can be used to 
produce the initial template.}
\rem{The downside of all these methods is both the requirement for a 
carefully modelled template, and that they do not generalize to character species 
with a significantly different anatomic structure from the given template.}

\subsection{Interactive volumetric and character modelling:} 
\citet{takayama2010volumetric} proposed a novel 
diffusion surface (DS) representation to model the smooth color variation seen in 
fruit and vegetables. 
User input to their approach, and others \cite{owada2008volume,pietroni2007texturing} 
is based on cross-sections, which are ill-suited to modelling complex muscle geometry 
and connectivity. Solid texture synthesis \cite{pietroni2010survey} is focused 
on modelling homogeneous material like wood or marble and \citet{cutler2002procedural} 
uses scripting to define internal volumetric structure of mesh(es). 
\citet{yuan2012object} do facilitate solid modelling of heterogeneous objects with 
multiple internal regions using multiphase implicit functions. 
However, these approaches are not artist-centric or require 
segmented and labeled 3D biomedical images as input.
 \citet{wang2011multiscale} represent complex internal 3D structure using multiscale vector 
volumes. The object is decomposed into components modelled 
as SDF trees. However, the user needs to actually create the “building-blocks” 
of an object, such as SDF instances and region definitions, and then assemble 
them together into linked SDF trees. 

Several sketch-based interfaces for character modelling  \cite{nealen2007fibermesh, 
takayama2013sketch, schmid2011overcoat, de2015secondskin} use a 3D geometric skin 
as a canvas on and around which to project 2D sketch strokes. Our work is similar 
in spirit to \cite{schmid2011overcoat,de2015secondskin} in that we are focused on 
drawing curves, around the skin, and specifically within a volumetric domain 
constrained by the surface of the skin. 

The actual shape of the muscles is inferred from the scalar values defined on the 
curves via a volumetric rendering approach. We do not explicitly generate the 
geometry (triangle or \tetl\ meshes) of the muscles until the user completes  the 
modelling session. Unlike previous methods, our approach allows the user to rapidly explore
and experiment with both the topological connectivity  and shape of musculoskeletal structures, with a guarantee of precisely constrained and intersection-free structures that conform outside-in to the given skin surface.

\section{Our system}
\begin{figure}
    \includegraphics[width=\columnwidth]{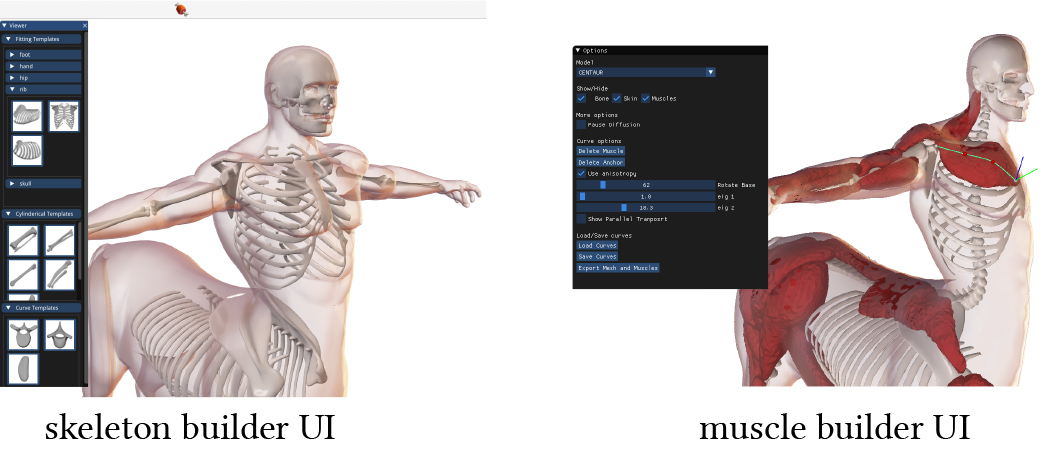}
    \caption{The skeleton and muscle builder UIs.} 
    \label{fig:ui}
\end{figure}

The user starts by providing the skin mesh of a model. The bone meshes can also 
either be provided by the user or built using our interface (Fig.\ref{fig:ui} left).
We generate the \tetl\ mesh $\T ∈ \R^{|\T| × 4}, \V \in \R^{|\V| × 3}$ 
from combined skin and bone geometries \rev{ to make sure our tetrahedralization 
conforms to the bone geometry}. \rem{The tetrahedrons that belong to the bones are 
given as an index vector into $\T$.} \rev{We remove all tetrahedra belonging to 
the bone geometry. This simultaneously induces desirable \emph{natural boundary conditions} (see, e.g.,
\cite{Stein2018}) and increases computational performance.}
A tablet or a mouse can be used to draw an open 2D stroke 
that starts and ends over the bone surface. The muscle surface corresponding 
to the drawn curve is presented to the user. The user can continue 
to draw more strokes or edit the existing ones to create a full muscular system 
(Fig.\ref{fig:ui} right). 
The results can then be exported as surface or \tetl\ meshes and \rem{used 
for animation, visualization, or} further edited in other software 
packages. In the following sections, we describe each part of our muscle 
modelling system and discrete implementation in detail. 

\subsection{Skeleton Authoring}
\label{section:skeleton_auhoring}
If pre-existing skeletal geometry is not available, we provide a tool for creation 
of the skeletal system from the pre-existing library of bones.
We define a three categories for the types for bones: cylinder bone, curve bone, 
shape bone. The user can select a pre-defined templates for each of the categories. 
Then each of template bones are properly placed inside the body mesh with the 
algorithm described below.

\subsubsection{Cylinder bone}

A cylinder bone is a type of bone that can be represented as a line segment 
(e.g. arm or leg bone). When a user clicks on a point on the surface of the 
skin mesh, we cast a ray through the mesh and record the first two hits, which 
correspond to the ray entering and exiting the mesh, respectively. We take the 
midpoint of these two intersections as one endpoint of the cylinder bone, and 
the other endpoint is determined interactively using the same ray-casting 
procedure while the mouse button is held down. A cylinder bone template is 
pre-rigged with two point handles and it is deformed accordingly as its two 
point handles are attached with the endpoints of the line the user draw.

\subsubsection{Curve bone}

Curve bone is a type of bone that can be represented as a curve. For example, a 
spine can be described as a sequence of vertebrae bones placed along a curve. 
When the user draws a curve, it is projected onto a user-defined plane of symmetry. 
Then for easy editing of the curve, we fit a Catmull-Rom spline to a given points 
on plane.  Finally, a user selected template is distributed along the spline.

\subsubsection{Shape bone}

A shape bone is a type that cannot be represented as a line segment nor a curve. 
Essentially, it describes all the bones with complex shapes. We place the shape 
bone as local fitting on the user-specified region. We determine the center of 
the template with a mouse click, using the same raycasting procedure used to 
compute the endpoints of a cylinder bone. Then from that initial shape, the 
template is iteratively fitted to its local region of the skin mesh.

\subsubsection{Local fitting} 

We pre-rigged the template with a cage deformer. Using data that already has 
skin and corresponding bone mesh, we define a cage by cutting a local region 
of the skin mesh with high decimation and with manual editing. 
We fit the cage using both step of rigid ICP~(Iterative Closest Point) and 
nonrigid ICP, and the templates are deformed using this registered cage. 
We first perform rigid ICP to find an optimal transformation of its closest 
corresponding target points. After it has converged within the threshold, we 
then perform nonrigid ICP by deforming the cage with an additional \rev{squared Laplacian smoothness} 
term to prevent abrupt deformation. Corresponding target points are 
determined by finding the closest points to cage vertices on the skin mesh. 
We discard the correspondence point whose normal directions are almost opposite 
to their closest projected points


\begin{figure}
    \includegraphics[width=\columnwidth]{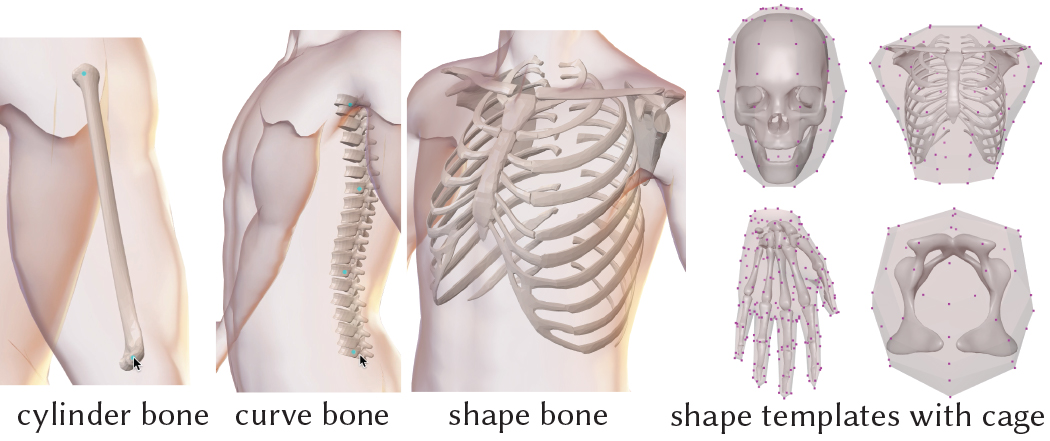}
    \caption{three types of bone in our skeleton authoring interface, and 
    pre-rigged on cage deformer of shape templates.} 
    \label{fig:diffusion}
\end{figure} 

\subsection{Muscle Curve Authoring}
\label{sec:curve authoring}

The user draws a curve for each muscle. Skeletal muscles require an origin and insertion points where the muscle 
tendons are being attached to the bone. In our interface we expect the user to 
always begin and end the stroke over the bone surface and provide the necessary 
visual feedback to achieve that. The first and last 
points of the curve are automatically projected via ray-casting onto the 
surface of the bone to find their corresponding 3d coordinates. 
\begin{wrapfigure}[11]{r}{0.23\linewidth}
    \vspace*{0.1\intextsep}
    \hspace*{-0.5\columnsep}
    \begin{minipage}[b]{\linewidth}
    \includegraphics[width=\linewidth, trim={0mm 4mm 0mm 0mm}]{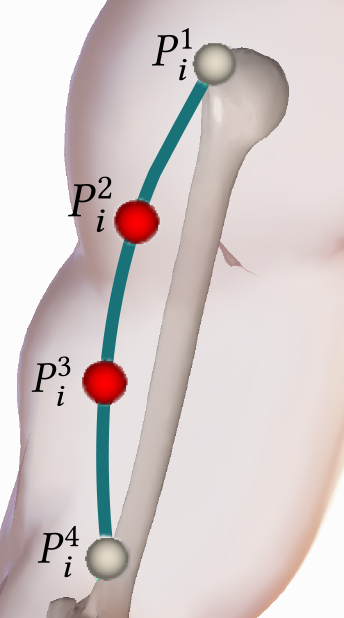}
    \label{fig:musclecurve}
    \end{minipage}
    \end{wrapfigure}
The depth value 
for all the points drawn in between is ambiguous, so we simply linearly 
interpolate the depths of the first and last points. 
Because we want the curves 
to be easily editable we fit a Catmull–Rom spline $\mathbf{S}_i$ in a least squares manner 
with 4 control points ${P_i^1, ..., P_i^4}$ by default: one for each end point 
and two along the curve (see inset). We denote the resulting muscle curve 
network as a set of splines $\{\mathbf{S}_1, \mathbf{S}_2, ..., \mathbf{S}_{m}\}$. 
Each control point ${P_i^p}$ is augmented with an additional attribute 
representing a \tissueit\ value ${D_i^p}$. Users can adjust the \tissue\ 
value to shape the muscle: a larger value results in a "thicker" shape 
around the control point. 
 
\subsection{Muscle and Fat Functions}
Let  $\Omega \in \R^3$ denote the volumetric domain defined by the \tetl\ 
mesh $\V$, $\T$. Our goal is to find a scalar muscle function $f_i : \Omega \rightarrow \R$ 
for each muscle $i$ that describes the likelihood of any point $p \in \Omega$ to 
belong to a muscle $i$. Similarly we define a fat function $f_s$ that describes 
the likelihood of any point to belong to a fat layer. We propose to define $f_i$ 
and $f_s$ as minimizers of the Dirichlet energy subject to constraints:
\begin{align}
    \argmin_{f_s,f_i, i=1,...,m} &\sum_{i=1}^{m} \int_{\Omega} \nabla f_i^T \A_i \nabla 
    f_i + \int_{\Omega} || \nabla f_s||^2  dV&   \label{eq:mfunc_energy} \\
    \text{subject to}  \quad &f_i|_{\delta\Omega_{i} = D_i} &\label{eq:mfunc_con_1} \\
    &f_i|_{\delta\Omega_{j} = 0} \quad \quad j \in \{1,...,m\},j \neq i &\label{eq:mfunc_con_2} \\
    &f_i|_{\delta\Omega = 0} & \label{eq:mfunc_con_3}\\
    &f_s|_{\delta\Omega = d_{\text{fat}}} &\label{eq:mfunc_con_4} \\
    &f_s|_{\delta\Omega_{j} = 0} \quad \quad j \in \{1,...,m\} \label{eq:mfunc_con_5}&
\end{align}
where $\A_i$  is an optional user-defined diffusion tensor field 
which biases the directions in which material flows at a point in space. 
Intuitively, we "diffuse" each muscle curve such that the \tissue\ values at 
the curve points ($\delta \Omega_i$) are set by the user (via interpolation of 
\tissue\ values $D^p_m$ at the control points) and the \tissue\ values at all 
other muscle curves ($\delta \Omega_j$) and skin ($\delta\Omega$) are set to zero. 
We additionally diffuse from the skin to represent the fat layer \rev{(fat function $f_s$, where "s" stands for skin)} by 
setting its \tissue\ value to a 
user-defined $d_{\text{fat}}$ and constraining the values at all muscle 
curves to be zero (Fig. \ref{fig:diffusion}). \rev{The fat function always diffuses isotropically
and hence \revtwo{is} written as a separate term.} \rev{The vertices that belong to the 
bone surface do not “diffuse” bone tissue material but instead participate in the 
optimization as natural (zero normal derivative) boundary conditions.} 

\begin{figure}
    \includegraphics[width=\columnwidth]{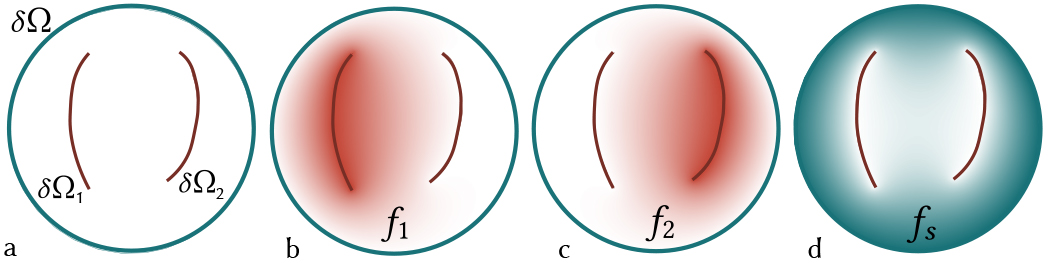}
    \caption{2D example of two muscle curves in red and skin layer in aquamarine (a).
    User defined tissue values at each curve ($\delta\Omega_1,\delta\Omega_2$) 
    are diffused (b,c)  to compute corresponding muscle functions ($f_1, f_2$). 
    Additionally fat tissue values defined at skin vertices are diffused to 
    compute the fat function $f_s$ (d).} 
    \label{fig:diffusion}
\end{figure} 

\subsection{Discretization}

The muscle splines usually do not coincide with the
vertices of the \tetl\ mesh and to discretize the splines we need to 
identify a set of \tets\ that contain each muscle curve $\it{\mathbf{S}_i}$. 
For each \tet\ $t$ and each muscle curve $i$ passing through it, we 
identify a point $\c_i^t \in \R^3$ on the curve that is contained inside the 
\tet\ (see inset). 

\begin{wrapfigure}{r}{0.15\textwidth} 
    \centering
    \vspace*{-1.3em}
    \includegraphics[width=0.15\textwidth]{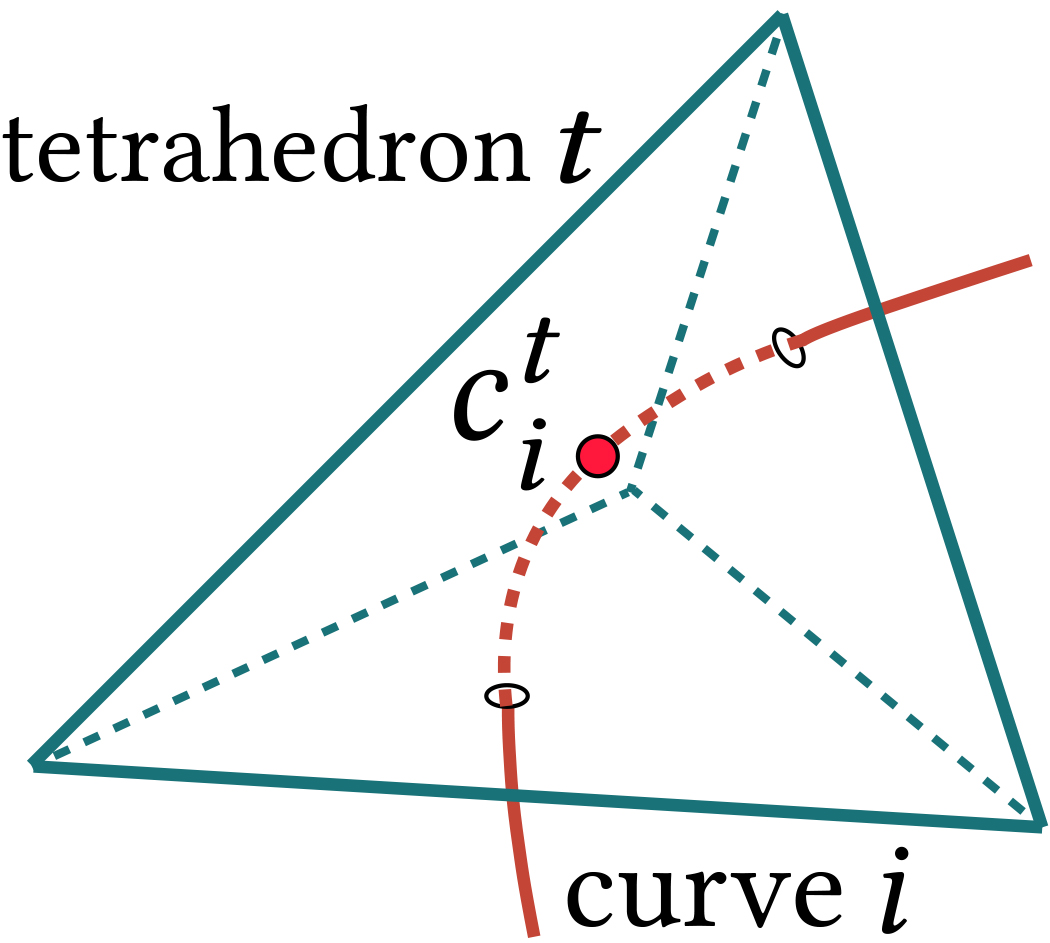}
    \hspace*{-1.3em}
    \vspace*{-1.3em}
    \label{fig:tetmidpoint}
\end{wrapfigure}

In practice, we choose the point to be the midpoint of the curve segment bounded 
by the \tet. We call this point a \textit{collocation} point. All 
collocation points can be found efficiently by finding the \tet\ containing 
the first curve point and then finding all the other \tets\ by tracing 
the curve, given that the adjacency is computed beforehand. 
This amounts to at most three ray triangles intersections per \tet\ and therefore 
very fast.

For each curve $i$, we stack all \textit{collocation} points in a matrix 
$\C_i \in \R^{|\C_i| \times 3}$ and compute a \tissue\ $d_i^t$ value for each 
\textit{collocation} point by interpolating values at control points 
$P_i^p$. One of the primary constraints to be satisfied 
(Eq. \ref{eq:mfunc_con_1}, \ref{eq:mfunc_con_2}) 
are the curve constraints, i.e.,the \tissue\ value at the vertices $\V$ 
must be determined to agree with the values at the collocation points. 
We use barycentric coordinates to interpolate the \tissue\ values for each \tet, 
such that the value $d_i^t$ of the collocation point $c_i^t$ is expressed as:
\begin{align}
    d_i^t = \sum_{j=1}^4 b_j * d_j
    \label{eq:tetmidpointbary}
\end{align}
where $d_j$ denotes the \tissue\ value of a vertex $j$ of a \tet\ $t$
(for $j=1,2,3,4$) to which the collocation point $c_i^t$ belongs to, and $b_j$ 
are the barycentric coordinates of $c_i^t$ with respect to the vertex $j$. 
Stacking the barycentric equation (\ref{eq:tetmidpointbary}) for a set of desired 
values of the collocation points into a matrix constitutes a linear 
equality constraint equation:
\begin{align}
    \label{eq:curve_constraint}
    \B_i\f_i = 
    \begin{bmatrix}
        \B_1\\\vdots\\\B_m
    \end{bmatrix} 
    \f_i=
    \begin{bmatrix}
        \d_1\\\vdots\\\d_m
    \end{bmatrix}=\d_i
    \in \R^{|\C|}
\end{align}
where for every curve $i=\{1,..., m\}$ we have 
$\B_i \in \R^{|\C| \times |\V|}$ which is a sparse matrix of stacked 
barycentric coordinates of collocation points for all curves with non-zero 
entry $\B_i(k,j)$ being a barycentric coordinate of the \textit{collocation} 
point $k$ with respect to vertex $j$, $\d_i \in \R^{|\C|}$ is a vector of 
stacked \tissue\ values of collocation points s.t. $\d_i(j)$ is nonzero only 
if collocation point $j$ belongs to curve $i$ and $\f_i \in R^{|\V|}$ are values 
of muscle function $f_i$ at each vertex of the \tetl\ mesh. The fat tissue constraint
(Eq. \ref{eq:mfunc_con_5}) can be similarly discretized as  
\begin{align}
    \label{eq:fat_constraint}
    \B_s\f_s = 
    \begin{bmatrix}
        \B_1\\\vdots\\\B_m
    \end{bmatrix} 
    \f_s= \mathbf{0}
    \in \R^{|\C|}
\end{align}
The Dirchlet Energy in Eq.\ref{eq:mfunc_energy} is discretized as 
\begin{align}
    \label{eq:energy_skin}
    \text{min}  & \sum_{i=1}^m (\f_i^T \tilde{\L}_i\f_i + \alpha ||\B_i \f_i - \d_i||^2) + \f_s^T\L_c\f_s + \alpha ||\B_s \f_s||^2\\
    \text{subject to} &\quad \f_i|_{skin} = \mathbf{0} \label{eq:energy_skin_1}\\ 
    &\quad \f_s|_{skin} = \d_{\text{fat}} \label{eq:energy_skin_2} \\ 
    &\quad \tilde{\L}_i = \G^T \tilde{\M} \A_i \G \label{eq:energy_skin_anisotropy}
\end{align}
where $\G$ is the gradient matrix (see \cite{botsch2010polygon} for derivation) 
and $\tilde{\M}$ is a mass matrix representing an 
inner-product accounting for the volume associated with each \tet, 
$\G^T\tilde{\M}\A\G$ is the anisotropic cotangent Laplacian 
(\cite{andreux2014anisotropic}), $\L_c$ is the standard (isotropic) cotangent 
Laplacian, $\alpha$ parameter that defines the tradeoff between smoothness of the 
resulting scalar field and respecting the \tissue\ values at the collocation 
points (set to 5 in our experiments). \rev{Because the collocation points are not
necessarily located at mesh vertices and many may appear in the same element, using soft constraints avoids overshooting.} Vertex values $\f_i$ of each muscle function 
can be computed separately and hence computation of Eq.\ref{eq:energy_skin} is 
easily parallelizable. \rev{Compared to regular-grid-based methods, the boundary-conforming tetrahedral mesh makes it easy to set precise boundary conditions.}

\subsection{Segmentation}
\begin{figure}
    \includegraphics[width=\columnwidth]{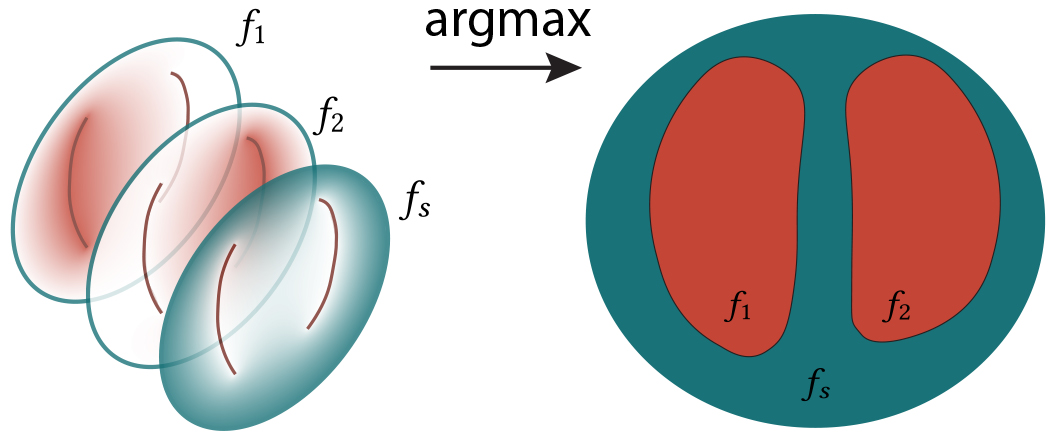}
    \caption{Each point in our domain has a vector of tissue values that represent 
    the likelihood of this point to belong to one of the tissues (left).
    We assign the point to the tissue with the highest probability (right).} 
    \label{fig:segmentation}
\end{figure}

Each point $p \in \Omega$ in our volume will either belong to a muscle, bone, or fat. The 
fat is visually represented as an "empty" space between muscles, skin, and bones. 
The points that belong to the bone tissue are simply all the points that are 
contained inside \tets\ comprising the bone geometry.
So we only need to differentiate between muscles and fat. We treat the muscle and 
fat functions as probabilities and assign the point $p$ to the tissue with 
the highest probability (Fig. \ref{fig:segmentation}):
\begin{align}
    \label{eq:softmax}
    tissue(p) = \argmax_i(\{f_1(p), ... f_m(p), f_s(p)\})
\end{align}
In the discrete case we find the \tet\ $t$ that contains the point $p$ and use barycentric 
interpolation to determine \tissue\ values for muscles and fat at $p$ given  
\tissue\ values ($\f_1^t..\f_m^t,\f_s^t$) at the vertices of the \tet\ $t$.
 \rev{We skip the normalization step of our probabilities because the muscle 
 visualization (Sec \ref{sec:visualisation}) and muscle extraction 
 (Sec \ref{sec:muscle_extraction}) are invariant to normalization.}

\subsection{Visualization}
\label{sec:visualisation}
In the end, we only want to visualize all points belonging to the muscles, while 
points belonging to fat should be invisible. A common solution to this problem 
in scientific visualization and computer graphics is volume rendering.
We take inspiration from the literature on volume rendering on 
unstructured grids (\cite{weiler2003hardware, silva2005survey}) which deals 
with rendering isosurfaces of a scalar function defined on vertices of the 
\tetl\ mesh. To achieve interactive rates we perform ray casting on the 
graphics hardware via a ray propagation approach and perform all computation inside 
a fragment shader.

To render the model we split each \tet\ into 4 triangles and submit them for 
rendering on the GPU. For each vertex of the triangle, we assign a vertex 
attribute with the value of the \tet\ index it belongs to and set it to not 
be interpolated when moving from vertex to fragment shader.
That way each fragment can be traced back to the \tet\ it belongs to.
We store all the information (vertex positions, normals, muscle and fat functions) 
about each \tet\ on the GPU and access it inside the shader.

\begin{wrapfigure}[11]{r}{0.20\linewidth}
    \vspace*{-0.5\intextsep}
    \hspace*{-0.5\columnsep}
    \begin{minipage}[b]{\linewidth}
    \includegraphics[width=\linewidth, trim={0mm 4mm 3mm -2mm}]{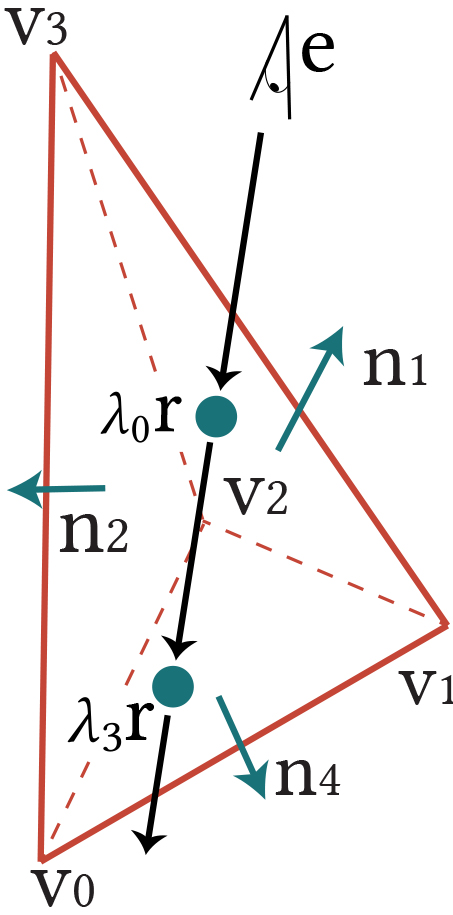}
    \label{fig:musclecurve}
    \end{minipage}
\end{wrapfigure}
Inside the fragment shader, we start with computing the entry point into the 
\tet\ that the current fragment belongs to by simply converting the 
fragment from screen space into camera space. We determine the corresponding 
exit point by computing three intersection points of the ray with the 
\rev{planes containing} faces of the entered \tet\ and choosing the intersection 
point that is closest to the eye point but not on a face that is visible from the eyepoint. 
With the eye point $e$, and the normalized direction $r$ of the viewing ray, 
the intersection points with the faces of the \tet\ are
$e+\lambda_ir$ with $0 \leq i <4$:
\begin{align}
    \label{eq:raytetintersection}
    \lambda_i = \frac{(v_{3-i} - e) \cdot n_i}{r \cdot n_i}
\end{align} 
where  $i \in \{0,1,2,3\}$ denote the face index, $v_i$ is the vertex opposite 
to the $i$-th face, $n_i$ is the normal vector of the face $i$ pointing outside 
of the \tet\ (see inset). 
A face is visible if the denominator in the previous equation is negative; 
thus, this test comes almost for free. If $\lambda_i$ is set to an appropriately
large number for all visible faces, $\min\{\lambda_i| 0 \leq i < 4\}$ identifies 
the exit point. Once the minimum $\lambda_i$ and its face $i$ are identified, 
the intersection point $x$ may be computed as $x = e + \lambda_i r$.

The muscle surface is only potentially visible if the entry point belongs to fat
in which case we ray march through the \rev{single} \tet\ from the entry point to 
the exit point until we \rem{hit a muscle tissue.} \rev{detect that the tissue 
changed from fat to muscle.} At which point we stop and compute the normal to 
shade the surface of the muscle. \rev{If the whole \tet\ belongs to fat we call the 
GLSL \texttt{discard} command and the GPU will call the shader again until we find a 
\tet\ that contains the surface of a muscle (if one exists).} 
Figures \ref{fig:teaser}, \ref{fig:anisotropy} show an example of a volume-rendered 
muscles.

\subsection{Anisotropy}
\label{sec:anisotropy}
\begin{figure}
    \includegraphics[width=\columnwidth]{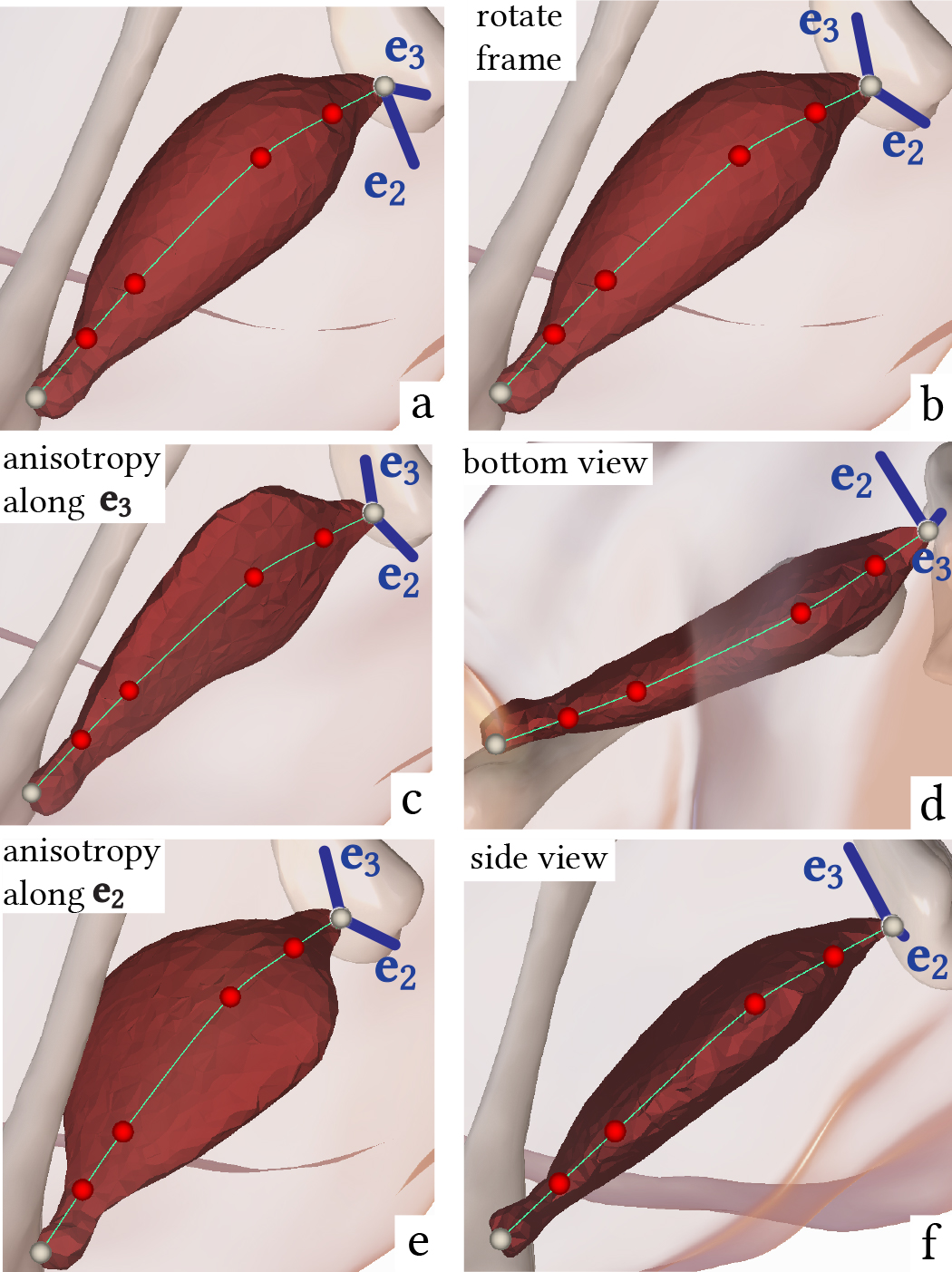}
    \caption{\rev{The user has additional control over the muscle shape 
 by changing the rate of diffusion along the certain directions.}} 
    \label{fig:anisotropy}
\end{figure}
To provide an additional control over the muscle shape we allow users 
to change the rate of diffusion along the certain directions.
This is achieved by introducing a tensor field that biases the directions in 
which tissues diffuse at a point in space. Tensor field is represented by a 
tensor matrix $\A_i \in \R^{|\G| \times |\G|}$ in Eq. \ref{eq:energy_skin_anisotropy}.

We provide an easy way for the user to construct the tensor fields for the 
individual muscles. 
The first point of each muscle curve is assigned a 3D frame 
$\Q = [\e_1 \e_2 \e_3]\in \R^{3 \times 3}$ 
representing eigenvectors of a tensor. The first vector in the frame $\e_1$ is 
always aligned with the curve tangent while the other two eigenvectors lie in 
its null space. The user has control over the rotation of $\e_2, \e_3$ along 
the axis defined by $\e_1$ (Fig. \ref{fig:anisotropy}ab). Additionally, the user can control 
the magnitude of eigenvalues $\lambda_2, \lambda_3$ to bias diffusion rate 
along $\e_2, \e_3$ directions (Fig. \ref{fig:anisotropy}cdef). We use the method of 
\cite{hanson1995parallel} to compute the frame $\Q_i^t$ at each collocation point 
$c_i^t$, and assign frames to each vertex of the \tets\ containing 
collocation points (using the frame of the closest collocation point).

Each frame $\Q$ can be considered as a set of nine scalars and the problem of 
propagating values from the fixed vertices of the \tets\ 
containing collocation points to the remaining vertices is thus same as 
computing nine scalar fields over the mesh. To find the \rev{unknown} values at the 
remaining vertices, we use the Laplacian smoothing framework, i.e. for each 
scalar field, we create a harmonic vertex-based scalar field on $\T$ given the 
boundary conditions (from the values at the fixed vertices) \cite{palacios2016tensor}. 
Once all 9 scalar fields have been obtained, we compute the per-vertex tensor 
$\Q^T \mathbf{\Lambda} \Q$  where $\mathbf{\Lambda}$ 
contains eigenvalues $\lambda_1 \lambda_2 \lambda_3$ along the diagonal and 
assemble the tensor field matrix $\A_i$ for the  muscle $i$.

\section{Muscle extraction}
\label{sec:muscle_extraction}
\begin{figure*}
\includegraphics[width=\textwidth]{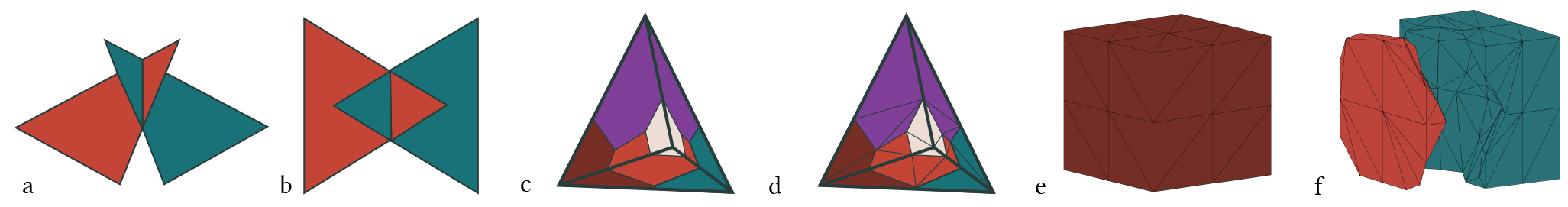}
\caption{ Two triangles (a) and the corresponding maximization diagram (b). 
Example of the maximization diagram cells of 5 different functions over the 
\tet\ (c) and tetrahedralization of each cell (d). Tetrahedral mesh (e) 
with 2 different functions defined over its vertices and the resulting tetrahedralized
maximization diagram (one of the parts is slightly moved in the figure to show 
the internal \rev{tessellation})} 
\label{fig:envelope}
\end{figure*} 
Given the complete muscle curve network and the corresponding muscle functions
defined over the volumetric domain, we need to extract the geometry of the muscles
suitable for downstream applications. 

\rev{Isosurface extraction methods \cite{lorensen1987marching, Labelle:2007:ISF, chentanez2009interactive} 
do not trivially solve our problem, because we are not simply extracting an isosurface 
of a scalar function. Instead, we want compute a \emph{pointwise} maximum of multiple 
scalar functions inside each tetrahedron and extract a boundary separating each 
function while ensuring the final output is a manifold tetrahedral mesh. 
In other words, inside each tetrahedron, each tissue function can be thought of 
as a point-wise vote for ownership.
We want to split the tetrahedron along boundaries that delineate changes in the
maximum vote and assign each sub-tet to the tissue with maximum value.} 
\begin{figure}
    \includegraphics[width=\columnwidth]{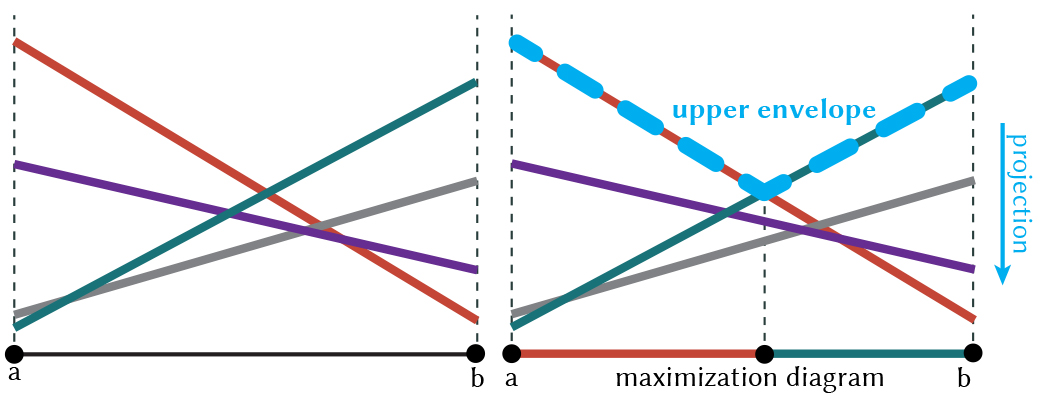}
    \caption{\rev{Four line segments (red, green, purple, gray) are defined over 
    a single 1D element $[a,b]$ (left). The upper envelope (light blue) and the 
    maximization diagram (right).} }
    \label{fig:line_upper_envelope}
\end{figure}
We cast the problem of extracting the muscle  
surface geometry as a solution to the computation of the upper envelope of scalar 
functions representing each tissue over each \tet. \rev{First, lets look at a simple 
example of the upper envelope problem in \reffig{line_upper_envelope}.
Four line segments are defined over the 1D domain $[a,b]$ 
(Fig. \ref{fig:line_upper_envelope}, left). The upper envelope 
is the point-wise maximum of all segments over the domain. The maximization 
diagram is a subdivision of the domain $[a,b]$ into cells, where each cell's
identity is induced by the upper envelope. Alternatively, we can think 
of the maximization diagram as a projection of the upper envelope onto the domain 
(Fig. \ref{fig:line_upper_envelope}, right).}

Let us now define the general 
upper envelope problem. Let $S=\{s_1, s_2.... , s_n\}$ be $n$ $d$-simplices in 
$(d+1)$-dimensional space. \rev{A $d$-simplex has ($d+1$) vertices, i.e. $d=1$ is 
a line segment, $d=2$ is a triangle and $d=3$ is a tetrahedron.}
We can thus view each $s_i$, as the graph of a partially defined linear function 
$x_{d+1}=f_i(x_1, x_2,... , x_d)$, whose domain of definition is a $d$-simplex, 
namely the orthogonal projection of $s_i$, onto the hyperplane $x_{d+1}=0$. The 
upper envelope, $M$, of the given simplices is the pointwise maximum of these 
functions \cite{edelsbrunner1989upper}, that is,
\begin{align}
    \label{eq:envelope}
    M(x_1,x_2,..., x_d) = \underset{ 1 \leq i \leq n}{\max} {f_i(x_1,x_2,...,x_d)}
\end{align} 
The maximization diagram $\it{M}_S$ of $S$ is  the subdivision of $R^{d}$ into 
connected cells obtained by the projection of the upper envelope of $S$ in 
the $x^d$ direction. \rev{The example in \reffig{line_upper_envelope} 
corresponds to $d = 1$ where each line segment is a 1D simplex in 2D space.}
\reffig{envelope}a shows an example of two 2-dimensional 
simplices in 3-dimensional space whose maximization diagram is shown in (Fig. \ref{fig:envelope}b). 
Solutions for solving the general upper envelope problem for $d=1$ and $d=2$ 
(\cite{agarwal1996overlay, meyerovitch2006robust}) and computing their corresponding 
maximization diagrams have been proposed. \citet{abdrashitov2019system} computes 
$d=2$ upper envelope of "part" functions to extract the smoothed
part boundaries.

We consider tissue functions 
$\mathcal{T} = \{f_1, ..., f_m, f_s \}$ (combination of muscle and fat functions)
defined as scalar fields over the vertices of our \tetl\ 
mesh and we notice that Equation \ref{eq:softmax} is the pointwise maximum 
(Eq.\ref{eq:envelope}) of tissue functions. In our problem ($d=3$) we are 
interested in finding maximization diagrams of all tissue functions over the 
volume constrained by the surface of the skin. This problem can be solved by 
considering finding the maximization diagrams of the tissue functions over each 
\tet. In which case in contrast to the general 
upper envelope problem where our domain is a continuous hyperplane, we restrict 
our domain by a $3$-simplex on that hyperplane. \reffig{envelope}c shows
an example of the maximization diagram of the functions defined by the scalar values 
at the vertices of the \tet\ and (Fig.\ref{fig:envelope}d) shows 
the tetrahedralization of each cell. As a result, the \tet\ is split into 
more sub-tets where each sub-tet has only one function that is the maximum. 
In other words, we can "assign" the sub-tet to one of the tissues. 
Performing this operation for every single \tet\ results in a new 
\tetl\ mesh where each \tet\ is assigned to one tissue. 
We can extract individual tissue shapes by simply combining all \tets\ 
that are assigned to that tissue. \reffig{envelope}e shows an example of 
taking a \rev{uniform} 3x3 \tetl\ 
mesh with two scalar functions defined over it and splitting it into two 
\tetl\ meshes (Fig.\ref{fig:envelope}f). The split is the result of 
computing tetrahedralized maximization diagrams of every \tet. 
\begin{figure}
\includegraphics[width=\columnwidth]{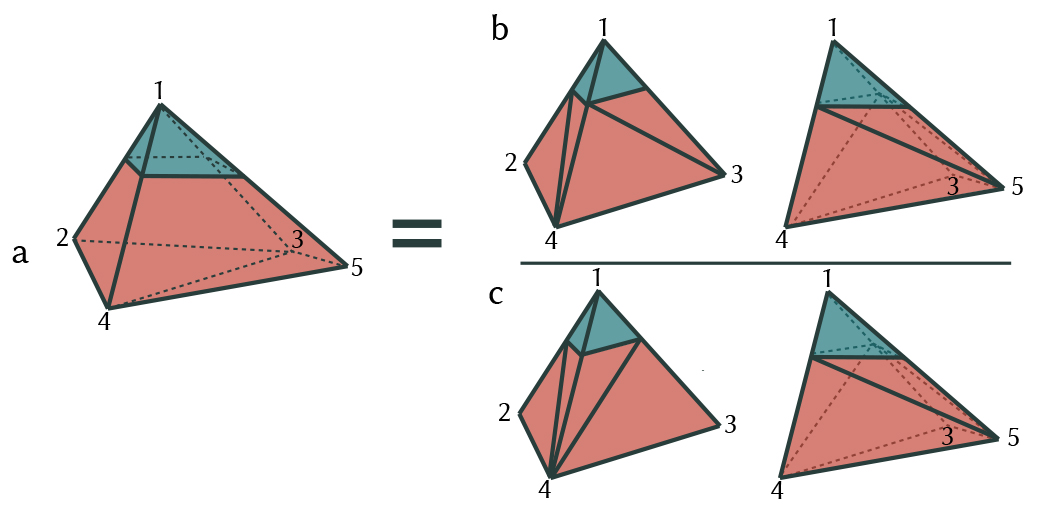}
\caption{Maximization diagram is defined over two adjacent \tets\ \textit{1234} 
and \textit{1345} (a). During the \rev{tessellation} there is no guarantee that 
the shared face \textit{\rev{134}} is split the same way (b). By making sure the 
order of edge splits is consistent we can guarantee that the face is split 
the same way (c).} 
\label{fig:tesselation_inconsistent}
\end{figure}
However, the resulting \tetl\ meshes representing each muscle are not 
guaranted to be manifold unless we are consistent with how we tessellate adjacent 
\tets\ (Fig. \ref{fig:tesselation_inconsistent}). In reality, we would 
just be getting a \tetl\ "soup" that is difficult to work with. Instead 
we want our resulting 
mesh to be manifold and hence we can easily use it for downstream \rem{geometry 
processing and simulation} applications. We propose an algorithm for computing 
manifold tetrahedralized maximization diagrams of functions defined as scalar 
fields over the vertices of a \tetl\ mesh. We first define auxiliary 
operations in Sections \ref{section:PruneTissues}, \ref{section:SplitTet} and 
then discuss the main algorithm in Sections \ref{section:CarveOut}.

\subsection{Prune tissues}
\label{section:PruneTissues}
We notice that if the tissue function $f_i \in \mathcal{T}$ is strictly below 
any other tissue 
function $f_j$, then it will not be part of the maximization diagram and 
hence can be ignored and help avoid unnecessary computations.
We define the sparse matrix $\W \in \{0,1\}^{m+1 \times m+1}$ where $\W_{ij}=1$ if 
and only if there exist at least one \tet\ in $\T$ whose maximization diagram 
contains cells assigned to tissues $i$ and $j$.

\subsection{Split tetrahedron}
\label{section:SplitTet}
\begin{figure}
    \includegraphics[width=\columnwidth]{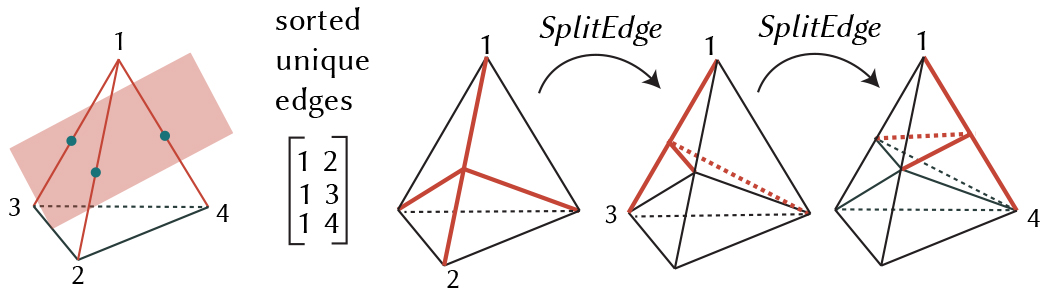}
    \caption{Split the edge over} 
    \label{fig:splitedge}
\end{figure}

Given only two different tissue functions $f_1$ and $f_2$
over a \tet\ we need to find the tetrahedralized maximization diagram that 
results from computing the upper envelope of those two functions. This problem 
is similar to the \textit{Marching tetrahedra} \cite{doi1991efficient} algorithm 
that finds an isosurface 
of a scalar field $f_1-f_2$ passing through isovalue of $0$. However, in 
our case we are not interested in extracting the isosurface but rather splitting 
the tet along the isosurface and tetrahedralizing the resulting polyhedrons.
\hfill \break \break
\textbf{Input}: \tet\ defined by 4 vertices $\V \in R^{4 \times 3}$, 
per-vertex values $\f \in R^4$ of a scalar function $f$.
\hfill \break 
\textbf{Output}: tet mesh $\T_{split}$,$\V_{split}$ resulted from splitting the input 
tet along the isosurface of $f$ at isovalue of $0$.
\hfill \break \break 
\textbf{Algorithm}: We find a set of \textbf{sorted} unique edges $\E$ of the 
\tet\ that 
are intersected by the isosurface of $f$. The number of edges $|E|$ is either 3 or 4. 
We then define a \textit{split\_edge} operator that given a \tet\ $t$ 
and an edge $e$ first splits the edge at a vertex $v$ and then splits $t$ 
into two \tets\ by connecting $v$ with two vertices of the edge opposite $e$ 
(Fig. \ref{fig:splitedge}). Similarly to \textit{Marching tetrahdera} the location 
of $v$ is determined by interpoloating vertices of $e$ using weights defined by $f$. 
We simpy run \textit{split\_edge} recusively on each $e \in \E$ to produce 
the resulting \tetl\ mesh $\T_{split}$. The order of edges in $\E$ determines 
the order in which we split the edges and 
hence determines the final topology of the \rev{tessellation}. By simply sorting 
the interested unique edges we solve the problem of inconsistent \rev{tessellation} 
(Fig.\ref{fig:tesselation_inconsistent}b) between adjacent \tets\ 
(Fig.\ref{fig:tesselation_inconsistent}c). 
We additionally maintain a history of unique edge splits before splitting 
an edge $e$ at vertex $v$. If the edge $e$  was split before at vertex $\hat{v}$ 
by a call to \textit{split\_edge} on one of the adjacent \tet\, we do not
create a new vertex $v$ but simply make it point to $\hat{v}$. To improve the 
robustness of the algorithm, when $v$ almost coincides with one of the 
vertices in $\V$ (when one of the values in $\f$ is close to zero within some 
thereshold) we simply remove that edge from $\E$ so it will not be split.  
\subsection{Tetrahedralized Maximization Diagram}
\label{section:CarveOut}
We iterate through every tissue function $f_i \in \mathcal{T}$ and use $\W$ to 
find a set of tissue funtions $\mathcal{\hat{T}}$ such that for any 
$f_j \in \mathcal{\hat{T}}, \W(i,j) > 0$. 
In other words $\mathcal{\hat{T}}$ contains all tissue functions that appear
in the maximization diagrams with $f_i$ in at least one \tet.
We compute all pairs $\{f_i,f_j\}$ of the tissue 
$f_i$ with tissues in the set $\mathcal{\hat{T}}$ . For every pair we 
iterate through all \tets\ $t \in \T$ with corresponding 4 vertices $\V_t$ and call 
the \textit{SplitTet} subroutine. Algorithm \ref{alg:all} summarizes our divide  
and conquer approach to computing tetrahedralized maximimzation diagram of 
tissue functions $\mathcal{\hat{T}}$ over the tetrahedral mesh $\T,\V$.
\begin{algorithm}
    \SetAlgoLined
    \KwIn{Tet mesh $\T$, $\V$ and per-vertex tissue values $\X \in \R^{|\V| \times (m + 1)}$}
    \KwOut{Tet mesh $\T_{out}$, $\V_{out}$ and per-tet tissue labeling $\L_{out}$}
    $\W$ = PruneTissues($X$); \# Section \ref{section:PruneTissues}
    \For{each tissue $f_i$ } 
    {
        $\mathcal{\hat{T}}$=FindIntersectingTissues($f_i$, $\W$)
        \For{every pair $(f_i, f_j) \in \mathcal{\hat{T}}$}  
        {    
            \For{$t \gets 1$ to $|\T|$}     
            {
                $\V_{st}$,$\T_{st}$ = SplitTet($\V_t$, $\X(t,i) - \X(t,j)$)\;
                //using barycentric interpolation to compute all tissues values 
                for all $v \in \V_{st}$ \\
                $\X_{st}$ = InterpolateDiffusion($\V_{st}$, $\T$, $\V$, $\X$)\;
                //replace $t$ with first \tet\ in $\T_{st}$, append the rest\\
                UpdateT($\T$, $\T_{st}$)\;
                $\V = [\V; \V_{st}]$\;
                $\X = [\X; \X_{st}]$\;
            }
        }
     }
     \caption{Tetrahedralized Maximization Diagram}
     \label{alg:all}
\end{algorithm}

\subsection{Fiber direction}


Fiber directions play an important role in simulation by defining the directions
the muscle contracts.  
Fibers tend to point in the same directions as gradients  of a hypothetical flow from one end of the muscle to the other 
\cite{choi2013skeletal, saito2015computational}. Given muscles \tetl\ mesh 
$\V_m, \T_m$, we leverage our muscle curve to guide the global flow of the fiber 
field. The flow should align with the tangent space of the muscle curve. This 
can be formulated via constraint that for each \tet\ along the curve we would like 
the projection of the gradient onto the normal space to be zero. 

Let $\T_{mc} \in \T_m$ be a subset of \tets\ that the muscle curve intersects. 
Then the desired "flow" can be found by solving the energy minimization problem:
\begin{align} 
\label{eq:fiberfield}
\min \u^T \L_c \u + \alpha(\N \G \u)^T (\N \G \u) \\
\text{subject to } \u_{start} = 0 \\
                \u_{end} = 1   \\
                \frac{\partial{\u_{skin}}}{\partial{\n}} = 0
\end{align}

\begin{wrapfigure}{r}{0.24\linewidth}
    \vspace*{-0.2\intextsep}
    \hspace*{-0.2\columnsep}
    \begin{minipage}[b]{\linewidth}
    \includegraphics[width=\linewidth, trim={0mm 10mm 0mm 2mm}]{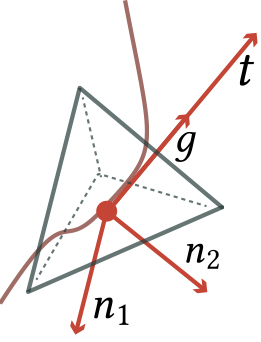}
    \label{fig:fiberdirection}
    \end{minipage}
\end{wrapfigure}
where $\G \in R^{3|\T_{mc}| \times |\V_m|}$ is a gradient matrix, 
$\N \in \R^{2|\T_{mc}| \times 3|\T_{mc}|}$ is sparse matrix containing 2 vectors 
per gradient vector \rev{representing}  
the null space of the curves tangent 
vector, $\u_{start}$ and $\u_{end}$ are a 
subset of vertices within a threshold distance of the curve endpoints.
This term 
forces the gradient of the scalar field in $\T_{mc}$ to align with the curve's 
tangent vector. The inset figure shows an example of a \tet\ that intersects 
the muscle curve where gradient $\g$ of the scalar field is aligned with the tangent 
vector $\t$ via constraining $\g$ to be orthogonal to the null space of $\t$ which is 
represented by $\n_1$ and $\n_2$.  
The fiber field can then be computed as the normalized gradient 
of $\u$, i.e., $\nabla \u/||\nabla \u||$.

\begin{figure}
    \includegraphics[width=\columnwidth]{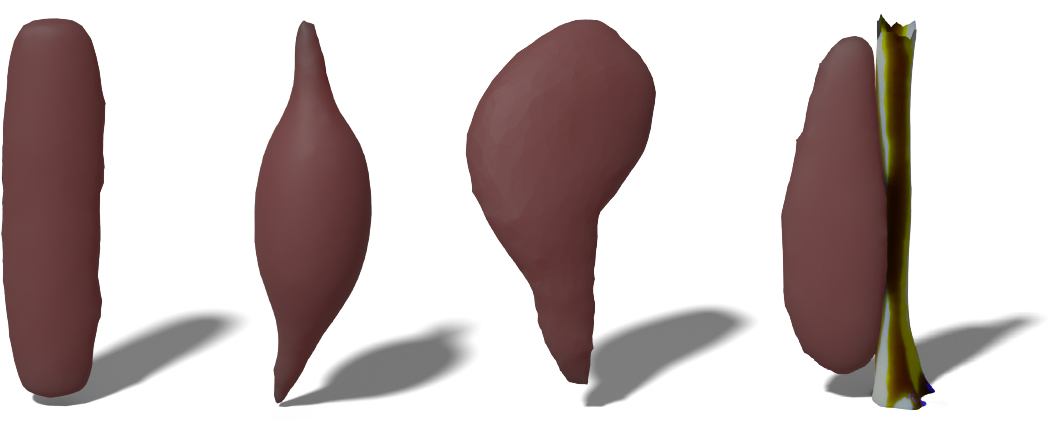}
    \caption{Parallel, Fusiform, Triangular, Unipennate. }
    \label{fig:muscletype}
\end{figure}

\section{Implementation and Results}
Skeletal and muscle curve authoring tools are implemented as 
separate standalone C++ applications using Eigen 
\cite{eigenweb}, libigl \cite{libigl}, and, for the user 
interface, ImGui. The output of the skeletal authoring tool are skin
and bone triangular meshes. We exploit the fact that due to symmetry one only 
needs to use half of the mesh for creating the muscles. Therefore, we cut both 
skin and bone meshes along the half-plane. The mesh vertices belonging to the 
cut boundary can be excluded from the constraints in Eq. \ref{eq:energy_skin_1},
\ref{eq:energy_skin_2} to help create muscles that are suppose to cross the cut boundary.
The half-skin and half-bone meshes are loaded by the muscle-curve authoring tool 
and we use TetGen \cite{si2015tetgen} to generate a \tetl\ mesh. 
The quadratic solver is implemented using the modified version of 
libIGL's \textit{min\_quad\_with\_fixed} function with 
Eigen's PardisoSupport module which provides significant speed ups. 
The tool was tested on a computer with Intel Xeon CPU @ 2.40GHZ, Nvidia GTX1080 
and 64GB of RAM and can perform at interactive rates for \tetl\ 
meshes with > 500k \tets. The geometry stays intact between curve editing 
operations, and we only update GPU buffers containing per-vertex diffusion values 
which is fast. 

\begin{figure*}
\includegraphics[width=0.91\textwidth]{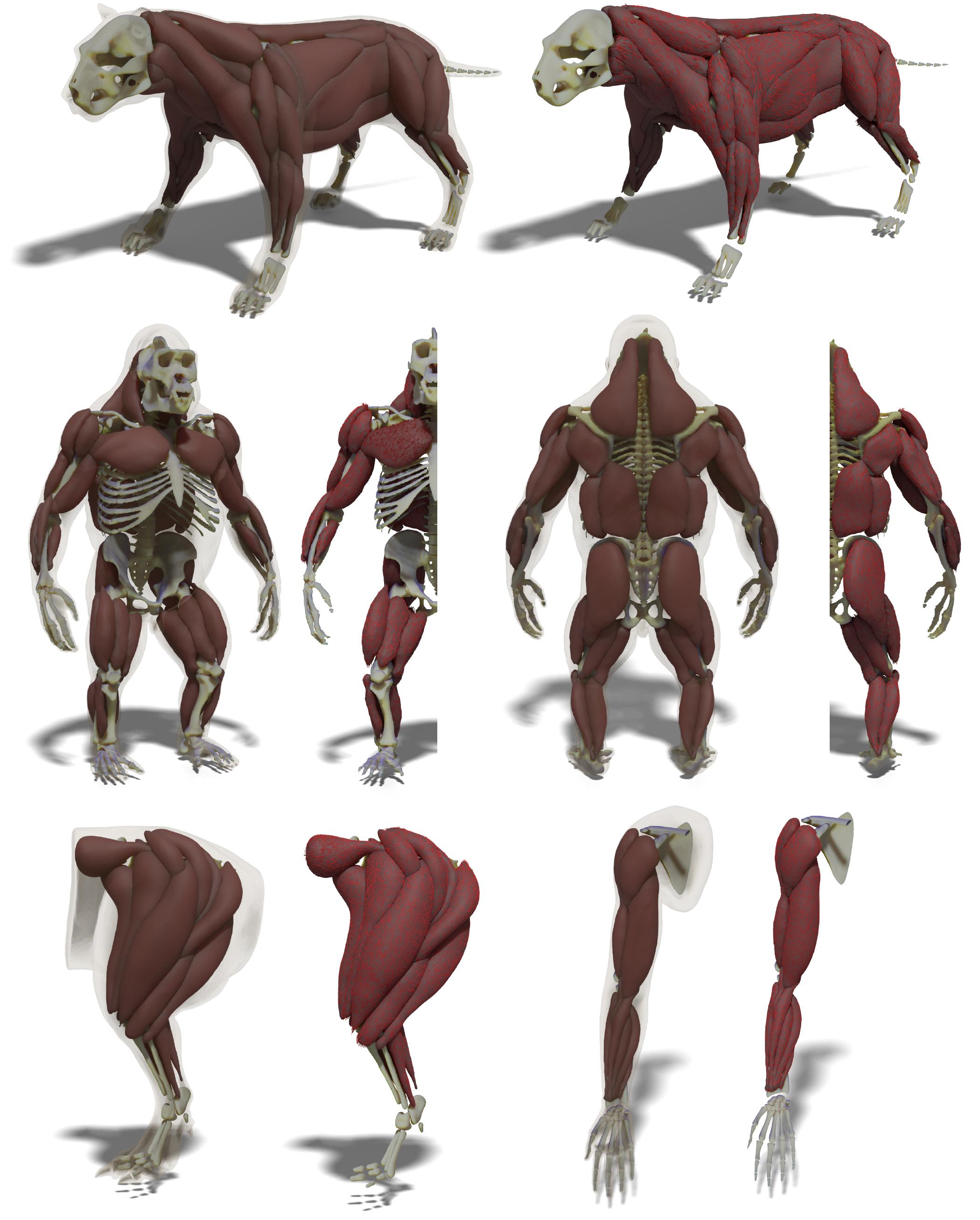}
\caption{Results. We refer the reader to the provided supplemental video to see 
the modelling process of the Ape model. We refer the reader to the provided 
\textit{Additional Modelling and Simulation Demos} video to see the modelling 
process of the Lion model. \rev{Lion (top) and Dinosaur Leg (bottom left) are provided 
by \copyright SideFx Software. Used under permission. Ape model (middle) is obtained from 
https://www.3dscanstore.com/ecorche-3d-models/gorilla-ecorche-3d-model. Used under permission.
Bone geometry of the arm (bottom right) is provided by 
\copyright Ziva Dynamics. Used under permission.}} 
\label{fig:results_authors}
\end{figure*}

Figure \ref{fig:results_authors} showcases the results created by the authors 
using our tool. We informally evaluated our tool with a professional animator who 
has experience with using Maya Muscles \cite{mayamuscle} for creating musculoskeletal 
systems. They mentioned that using existing tools requires a significant amount of 
tedious work for ensuring that muscle geometry respects the geometry of its 
surroundings (bones, skin, other muscles), and the overall workflow is "unintuitive" and "awkward". 
They liked the fact that our tool allows one to simply draw the location of the muscle while
handling these issues automatically and provides enough intuitive controls 
(changing tissue values, moving control points, enabling anisotropy) 
to create complex muscle shape (Fig.\ref{fig:muscletype}). 
One feature they wanted to see added is volume preservation of the muscle: 
when a muscle is being flattened in one direction, there is a corresponding 
stretching in another direction so that the volume stays the same. They also 
wanted an ability to specify multiple attachment points instead of just two 
to enable finer control, especially for triangular types of muscle shapes.
We further informally evaluated our tool with an orthopedic surgeon with  
extensive knowledge of human anatomy but who does not have experience with 
3D modelling applications. They see the potential of using this tool to explain 
the surgery process to their patients. By using a pre-existing anatomical muscle 
model made with our tool, they would be able to edit the muscles to explain their 
medical procedure. However, similar to the 
animator, they mentioned that to create truly anatomically correct muscles, they 
need to be able to specify multiple attachment points.
 
\begin{figure}
    \includegraphics[width=\columnwidth]{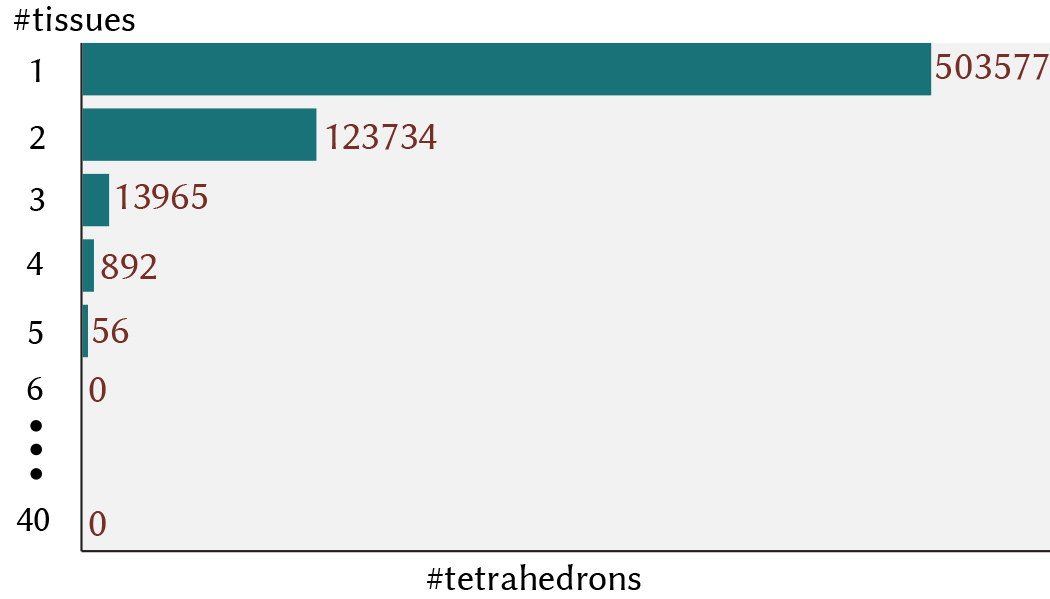}
    \caption{We consider the Lion model in Fig. \ref{fig:results_authors} with 39 muscles 
    and 642224 \tets. The plot shows the number of \tets\ and the maximum number 
    of tissue functions that can appear in their corresponding maximization diagram. 
    This shows that for the vast majority of \tets\ (78\%) computing their tetrahedralized 
    maximization diagrams requires no splitting or only trivial non-recursive 
    splitting of 2 tissues (19\%). No \tets\ require resolving intersection of 6 
    or more tissues.} 
    \label{fig:tissue_bar_plot} 
\end{figure} 

The muscle extraction algorithm in Section \ref{sec:muscle_extraction} uses the 
divide and conquer approach by considering pairs of tissue functions. In 
Figure \ref{fig:tissue_bar_plot}, using the example of the Lion model (Fig.\ref{fig:results_authors}),
we show that in practice the vast majority of \tets\ only contain 1 tissue 
function in their maximization diagram and therefore are not split. While 
the vast majority of \tets\ that do need to be split only contain 2 tissue functions, 
which means they are only split once. 

\begin{figure}
    \includegraphics[width=\columnwidth]{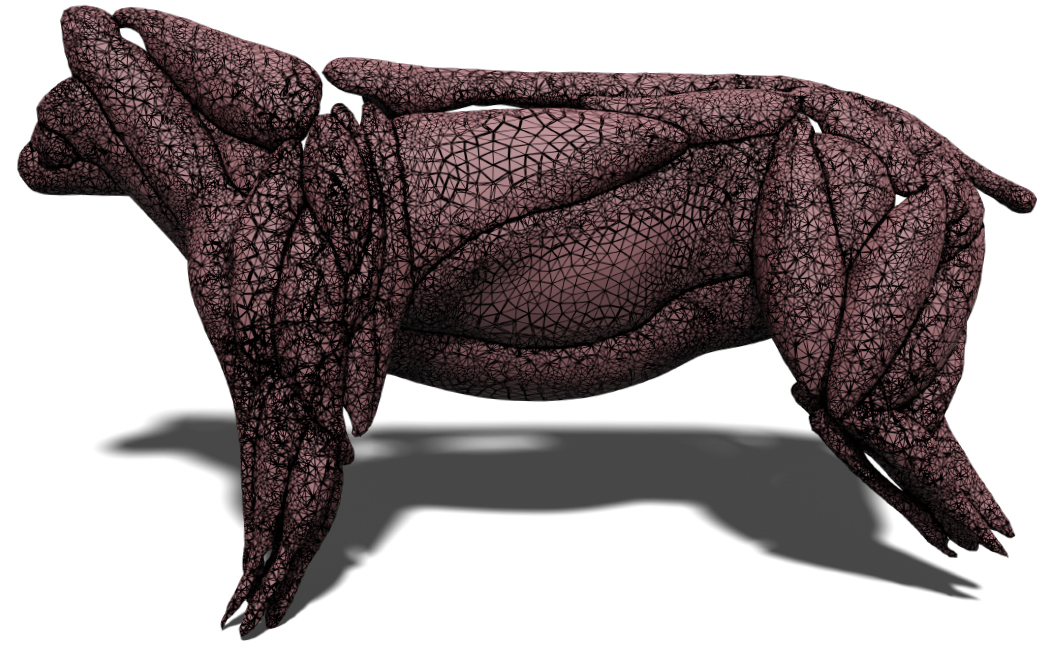}
    \caption{Example of the final \rev{tessellation} produced by the upper \rev{envelope} algorithm.} 
    \label{fig:upperenvelope_wireframe}
\end{figure}  

Because our volumetric domain is represented as an unstructured mesh, the 
resulting topology of the muscle meshes ends up containing \textit{skinny} 
\tets\ as shown in Figure \ref{fig:upperenvelope_wireframe}.
However, because we guarantee that the resulting meshes are manifold, it is 
straightforward to extract the boundary faces (faces that belong to only one 
tetrahedron). The resulting muscle surface triangular meshes can be easily 
remeshed and tetrahedralized again.
In practice, we simply run TetWild \cite{hu2018tetrahedral} with default 
parameters on the surface mesh to improve the topology. \rem{The tetrahedralized muscle 
geometry can be directly used for downstream tasks like simulation} 
\rem{and the surface geometry can be used for visualization.}  
\begin{figure}
    \includegraphics[width=\columnwidth]{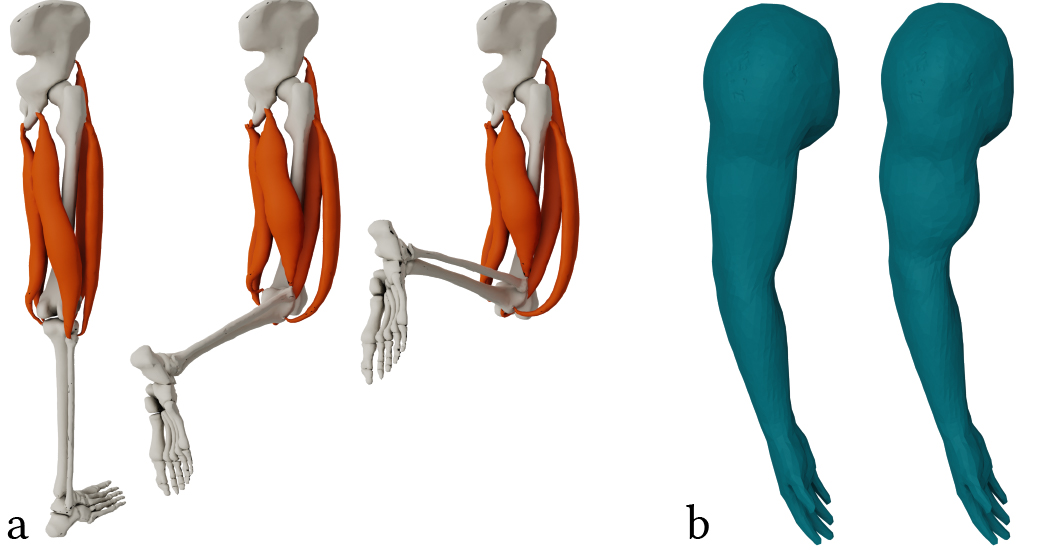}
    \caption{Simulating the motion of a leg using muscle geometry created with 
    our tool \rev{without contact handling}. Motion is induced by contracting
    the hamstring using the method of \citet{modi2020emu} (a). Dynamic Finite 
    Element simulation of isometric contraction of the bicep 
    using linear tetrahedral finite elements. Muscles are modeled as Neohookean 
    elastic solids using the \rem{orthotropic} fiber model of 
    \citet{10.5555/846276.846285}. 
    Time integration is performed via Implicit Euler time stepping using the 
    Bartels library~\cite{bartels}. We refer the reader to the provided 
    \textit{Additional Modelling and Simulation Demos} (07:30m) video to see 
    the animated examples. Bone geometry (left) is provided by \copyright 
    Ziva Dynamics. Used under permission.}
    \label{fig:simulation_leg}
\end{figure}   
 
\rev{During the modelling process the muscles seen by the user are directly volume rendered 
upper-envelopes: rendering does not require performing any meshing operations 
during the modeling process. Every time the user creates or edits 
muscle curves we recompute values of the tissue functions (Eq. \ref{eq:energy_skin}) for each 
vertex of the tetrahedral mesh and update those values in the corresponding 
GPU buffer. These values are later accessed by the fragment shader to perform 
the volumetric rendering using the ray-marching method (supported by all modern 
GPUs). The muscle visualization using fragment shaders is fast (>30fps) on our system.
The bottleneck of our system during modeling is computing the solution to (Eq. \ref{eq:energy_skin}) 
for which we currently use  the Eigen library with Pardiso enabled. This solve 
immediately benefits from any generic Poisson solving optimizations which are 
orthogonal to our main contributions. We only perform  a meshing operation 
(Sec. \ref{sec:muscle_extraction}) after the modelling session is complete, to 
export muscles as tetrahedral meshes for downstream applications. Using muscle 
extraction every time we create/edit a muscle would negatively impact performance, 
which is the motivation for our volumetric rendering approach. }


\section{Limitations and Future Work}
Although we demonstrated that our tool enables users to create complex 
muscular-skeletal geometries, there are limitations, subject to future work.

The quality of the muscle shapes depends on the quality of the 
tetrahedrazation ($\V$,$\T$) of the input skin and bone meshes.
On one hand, coarse meshes prevent the creation of smooth and thin muscles; on 
the other hand, very high quality meshes can affect 
the performance of computing muscle functions and hinder the interactive experience. 
This makes it hard to create muscles in the areas where bones are too close 
to the skin, unless the area is densely tessellated. We also notice that often a 
large portion of the model ends up being unused (the intestines area), yet it's 
still being used for computation of muscle functions which wastes computation.  
In practice, for all models, we were able to find a balance between the 
quality of the tetrahedralization and performance. In the future, we want to explore  
adaptive boundary \revtwo{conforming} \rev{tessellation} of the volume with 
increased \revtwo{tessellation}
quality only around boundaries of the muscles.

We use a simple heuristic (Sec. \ref{sec:curve authoring}) for 
creating initial muscle curves, which leads to multiple editing operations of 
control points to get the muscle curve shape just right. This makes it challenging 
to create muscle curves around complex bone geometry, since control points 
must be edited from multiple viewpoints. In the future, we would like to explore 
a more robust curve creation techniques that simultaneously utilizes skin and 
bone geometry to create the best guess for the initial shape of the muscle curve.

Currently, each muscle is represented by a single curve. Based on the feedback 
we received from users, we plan to explore how we can use multiple curves to 
represent a single muscle. This would allow us to create more anatomically 
correct and complex muscle shapes with multiple origin and insertion points. 
\rev{Additionally, there are many muscles in which fiber direction does not match the 
flow between origin and insertion and therefore we plan to explore ways to 
specify the pennation angle (the angle between the longitudinal axis of 
the entire muscle and its fibers) as it is an important parameter in muscle 
contraction dynamics.}

\rev{We have investigated the potential use of the resulting muscle geometry for 
simulation (Fig. \ref{fig:simulation_leg}) but more work needs to be done to 
test simulations with the complex and densely packed full-body muscle systems. 
With the help of professional artists, we plan to evaluate how animation-ready 
our resulting muscle geometry is using industry-standard software like Houdini 
\cite{sidefx} and \citet{ziva}.}

\rev{Our system is designed with the idea of creating muscle geometry from scratch 
but we have experimented with extracting curve networks from pre-existing artist-made muscle shapes 
(using methods for computing curve skeletons of 3D shapes) and importing them into 
our system, which can provide a good starting point.}

\rev{We currently do not support locking of the muscle shape. However, we could 
extract the mesh of a muscle, and treat it similarly to bones so that other muscles 
wrap around it. This requires to redo the initial \revtwo{tessellation} to make sure it 
conforms to imported muscle geometries but TetGen only takes a few seconds. 
Similarly an artist can import existing muscle meshes and continue to create 
muscles around them.}

\begin{acks}
    Our research is funded in part by NSERC Discovery (RGPIN-2017-05524, RGPIN2017–05235, RGPAS–2017–507938), 
    NSERC Accelerator (RGPAS-2017- 507909), Connaught Fund (503114), CFI-JELF Fund,
    Canada Research Chairs Program, New Frontiers of Research Fund (NFRFE–201), 
    the Ontario Early Research Award program,  the Fields Centre for Quantitative 
    Analysis and Modelling and gifts by Adobe Systems, Autodesk and MESH Inc.
    
    We thank Sarah Kushner and Abhishek Madan for proofreading; Vismay Modi for 
    helping to generate the simulation results; Oded Stein for helping with figures; 
    anonymous reviewers for their helpful comments and suggestions. Special 
    thanks to SideFx software for providing their models.
\end{acks}

\bibliographystyle{ACM-Reference-Format}
\bibliography{musclegeometry-acmtog}

\end{document}